# YRu$_3$B$_2$ – a kagome lattice superconductor

Michał J. Winiarski[1,2*], Dominik Walczak[1,2], Szymon Królak[1,2], Duygu Yazici[1,3] Robert J. Cava[4] and Tomasz Klimczuk[1,2†]

[1] *Faculty of Applied Physics and Mathematics, Gdansk University of Technology, Narutowicza 11/12, 80-233 Gdansk, Poland*

[2] *Advanced Materials Center, Gdansk University of Technology, Narutowicza 11/12, 80-233 Gdansk, Poland*

[3] *The Scientific and Technological Research Council of Turkey, Atatürk Bulvarı No: 221, Kavaklıdere 06100, Ankara, Turkey*

[4] *Department of Chemistry, Princeton University, Princeton, New Jersey 08544, USA*

\* michal.winiarski@pg.edu.pl

† tomasz.klimczuk@pg.edu.pl

**Abstract**

We report the synthesis and physical properties of a polycrystalline, hexagonal boride YRu$_3$B$_2$. Our resistivity and heat capacity measurements indicate that YRu$_3$B$_2$ is a weakly coupled superconductor, with critical temperature $T_c$ = 0.63 K and upper critical field $\mu_0 H_{c2}(0) = 0.11$ T. Density functional theory calculations, together with chemical-bonding analysis, reveal that the electronic states at and near the Fermi energy level are dominated by the Ru kagome sublattice.

**Introduction**

The kagome lattice is a canonical platform for exploring how electronic topology shapes physical properties of solid-state compounds[1–8]. Within the simplest nearest-neighbor tight-binding model, the band structure of the kagome lattice hosts a topologically flat band (TFB), generated by destructive interference of wavefunctions and the resulting formation of compact localized states[9,10]. Moreover, the hexagonal symmetry of the lattice results in a protected band crossing at the K point of the first Brillouin Zone (BZ) - the Dirac point. Thus, within a single material, tuning the Fermi level can drive a transition from a strongly correlated metal, when it lies within the TFB, to a Dirac semimetal when it approaches the Dirac point. It is this diversity and tunability between different ground states that kagome materials have attracted considerable research interest in recent years [10–14]. In real materials, however, the kagome network is never fully isolated from the neighboring atoms, resulting in the TFB attaining some dispersion. In a recent paper, Jovanovic and Schoop discussed the relations between the occurrence of the kagome bands and details of crystal structures in kagome network-bearing compounds [10].

Among the many distinct ground states possible to realize in kagome network materials, superconductivity has emerged as one of the most actively explored. In this context, LaRu$_3$Si$_2$ serves as a relevant example. It is isostructural to YRu$_3$B$_2$ at temperatures above 600 K, below which it undergoes an orthorhombic distortion[15]. In addition to a relatively high critical temperature, T$_c$ ≈ 7 K, LaRu$_3$Si$_2$ hosts multiple charge-ordered states and exhibits a dome-like shaped pressure dependence of the critical temperature T$_c$(p) [16]. These features suggest an interplay between charge order and superconductivity, possibly of unconventional character. Studies of structurally related compounds are therefore of great interest, as they can help determine whether the interplay between charge order and superconductivity is a general feature of kagome network superconductors, and thus an intrinsic consequence of the crystal lattice topology, independent of the specific elemental composition.

YRu$_3$B$_2$ crystallizes in the CeCo$_3$B$_2$-type structure, in which the kagome network is formed by Co atoms [17,18]. Honeycomb boron networks cap the Co$_3$ triangles and connect the stacked kagome layers (Fig. 1(a,b)), with the large Ce atoms above and below the Co-hexagon centers. In this family of intermetallic compounds, superconductivity was reported for ThRu$_3$B$_2$ (T$_c$ = 1.6-1.8 K) and (Y$_{0.5}$Th$_{0.5}$)Ru$_3$B$_2$ (T$_c$ = 1.4-1.5 K) [17]. Recently, a machine learning study predicted a superconducting critical temperature T$_c$ = 2.121 K for YRu$_3$B$_2$ [19].

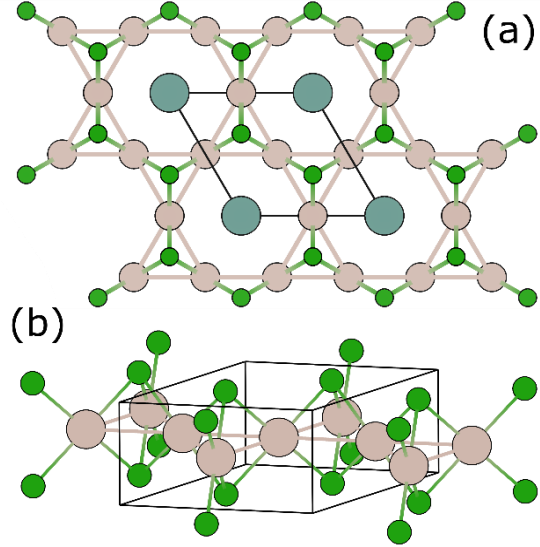

Fig. 1 Panel (a) Crystal structure of YRu$_3$B$_2$, viewed along the *c* direction (Y atoms are shown in grey, Ru – beige, B – green). Panel (b) shows the Ru kagome network capped by two B layers.

In this work, motivated by theoretical predictions and the kagome structure of the Ru sublattice, we have synthesized and studied polycrystalline samples of the hexagonal YRu$_3$B$_2$ boride. Our electrical resistivity and heat capacity measurements reveal that YRu$_3$B$_2$ is a weakly-coupled superconductor, with the critical temperature T$_c$ = 0.63 K. In addition we performed density functional theory (DFT) calculations revealing that the electronic states around the Fermi level are mostly derived from the Ru 4*d* atomic orbitals that form the kagome sublattice. Although a topological flat band is not observed, the Fermi level crosses several weakly-dispersive bands, and characteristic band crossings at the K point are observed.

**Materials and Methods**

A polycrystalline sample of the YRu$_3$B$_2$ was prepared using a two-step arc melting technique. The synthesis procedure was performed in an inert, Zr-gettered high-purity Ar atmosphere using an MAM-1 GmbH Edmund Bühler arc furnace. First, YB$_2$ was made by carefully melting the appropriate amounts of high-purity yttrium (99.9%, Ames Laboratory) and boron (99.9%, Alfa Aesar). This step was challenging because elemental boron often scatter out upon initial contact with the arc, resulting in boron-deficient products. Only ingots exhibiting negligible mass loss were selected for further synthesis. During the second melting, YB$_2$ was reacted with metallic Ru. To enhance homogeneity, the resulting ingot was inverted and re-melted multiple times (4-5 minimum).

Phase purity was checked using the powder X-ray diffraction (pXRD) method with a Bruker D2 Phaser diffractometer equipped with a XE-T detector (Cu-Kα radiation). The structural parameters of the studied sample were obtained by the Le Bail analysis of the collected pXRD data, which was carried out with Bruker Topas software. The chemical composition of the YRu$_3$B$_2$ sample was examined using energy-dispersive X-ray spectroscopy (EDX) with a FEI Quanta FEG 250 scanning electron microscope. Temperature-dependent heat capacity and resistivity measurements were performed in a Dynacool PPMS system (Quantum Design) equipped with a $^3$He setup. Heat capacity data were collected using a standard thermal relaxation technique. Electrical resistivity measurements were performed using a four-probe technique with an applied current of 5 mA (1.8 K < T < 300 K) and 0.5 mA (0.5 K < T < 1.8 K). Electrical contacts were made by spot-welding platinum wires (ϕ = 50 μm) onto the polished surface of the samples.

Electronic structure calculations were performed within density functional theory (DFT) using the Quantum Espresso 7.2 package [20–22]. The projector-augmented-wave (PAW) [23,24] method was employed together with the Perdew–Burke–Ernzerhof (PBE) generalized-gradient approximation for the exchange–correlation functional [25]. A 7x7x13 *k*-point mesh was used for the self-consistent field (SCF) run, while for non-SCF calculations of the density of states and Fermi surface, the grid density was increased to 21x21x38.

Crystal Orbital Hamilton Population analysis was performed using the LOBSTER 5.1.1 code [26,27] by projecting the results of the plane wave calculations to a local orbital basis set by Bunge et al. [28].

The Fermi surface was visualized using the FermiSurfer 2.4.0 program [29]. Crystal structures were rendered using VESTA 3.4.5 [30].

**Results**

Figure 2 presents the results of the Le Bail refinement of the room temperature powder X-ray diffraction (pXRD) data for $YRu_3B_2$. All pXRD reflections were indexed within the hexagonal space group *P*6/*mmm* (s.g. # 191) with the lattice parameters: *a* = 5.4826(2) Å and *c* = 3.02166(1) Å, consistent with the values previously reported by Hiebl, et al. [19].

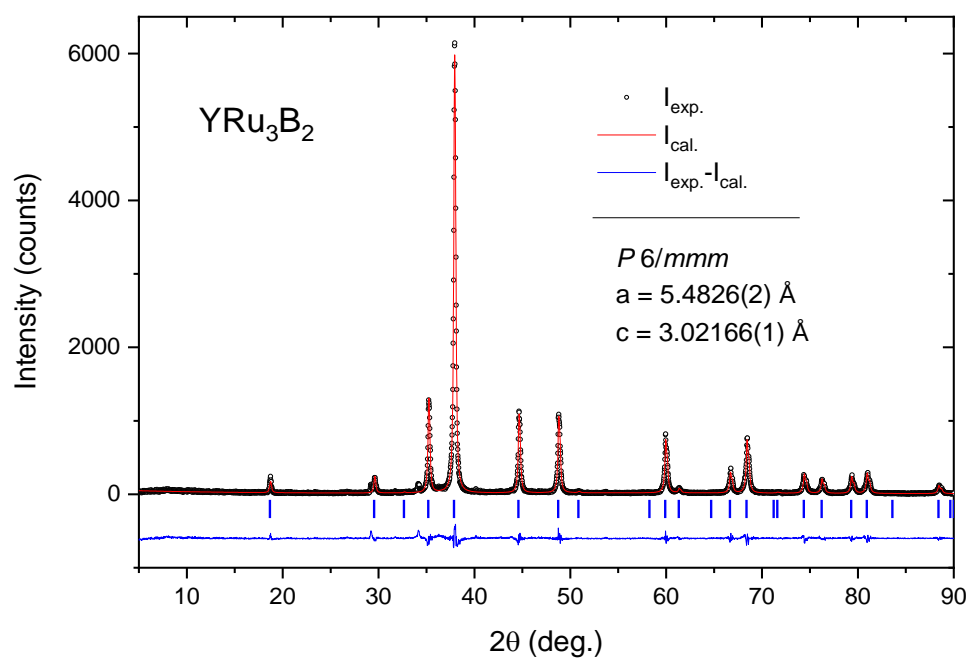

Fig. 2 pXRD pattern of $YRu_3B_2$ along the LeBail refinement. The black dots represent the experimentally observed data, and the red line represents the calculated pXRD pattern using the *P*6/*mmm* crystal structure. The blue line and vertical bars represent the difference in intensity between the experimental and calculated data and the expected Bragg reflections, respectively.

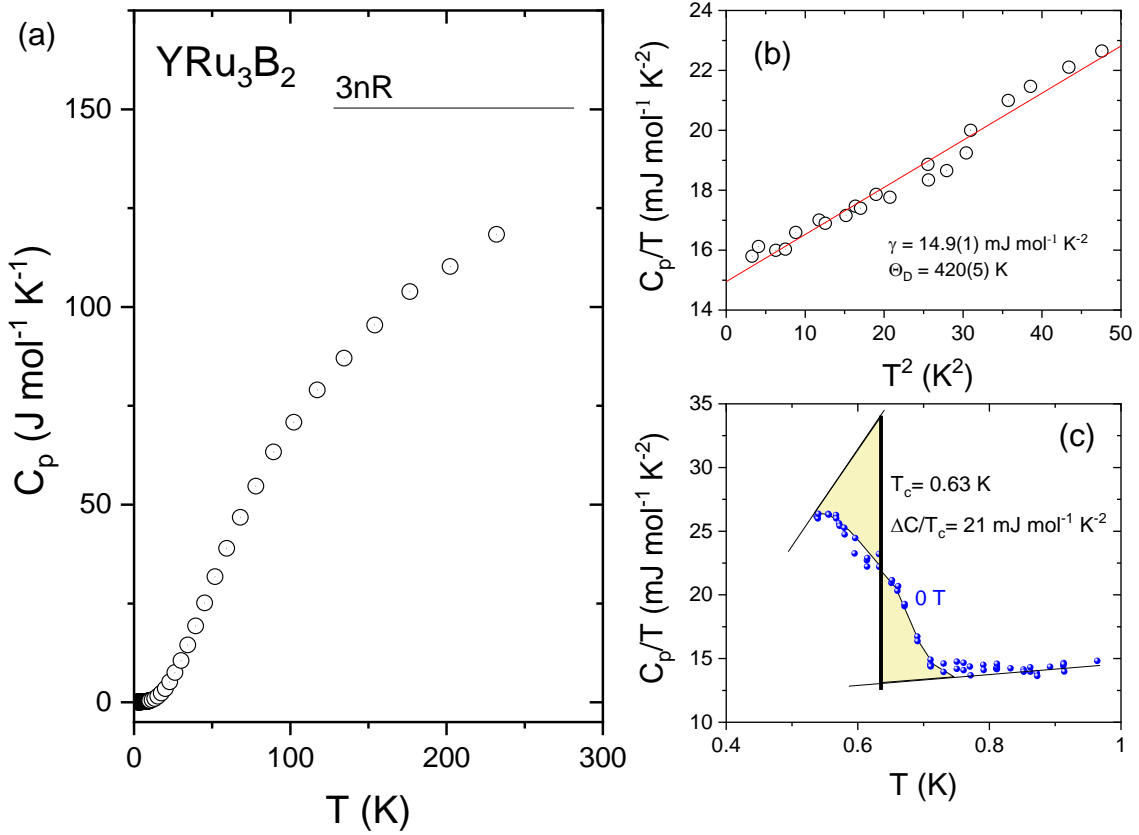

Fig. 3 Heat capacity of YRu$_3$B$_2$. Panel (a) shows C$_p$(T) plot in the 2-230 K temperature range. Panel (b) shows a linear fit to the C$_p$/T vs. T$^2$ (for T > T$_c$) data used to obtain the values of the Sommerfeld coefficient and the Debye temperature. Panel (c) shows the heat capacity around the superconducting transition temperature with a clear jump, suggesting its bulk nature.

Following the structural analysis, we investigated the heat capacity of YRu$_3$B$_2$. Figure 3 presents the heat capacity data acquired in the $^4$He system (panels a and b) and the low temperature data (panel c) obtained with the $^3$He system. As shown in panel (a), no phase transition is evident at elevated temperatures. The specific heat capacity (C$_p$) at 230 K is approximately 80% of the expected value derived from the Dulong-Petit law, 3nR = 150 mJ mol$^{-1}$ K$^{-2}$. This indicates a relatively high value for the Debye temperature. The Sommerfeld coefficient ($\gamma$) and Debye temperature ($\Theta_D$) estimates were obtained by fitting the low-temperature C/T versus T$^2$ (panel (b)) to the formula $C/T = \gamma + \beta T^2$. In this formula $\gamma T$ and $\beta T^3$ are the electronic and the phonon (Debye) contribution to the specific heat. The fit yields $\gamma$ = 14.9 mJ mol$^{-1}$ K$^{-2}$ and $\beta$ = 0.157(6) mJ mol$^{-1}$ K$^{-4}$, from which the Debye temperature was calculated to be 420(5) K using the following equation:

$$\theta_D = \left(\frac{12\pi^4}{5} nR\right)^{\frac{1}{3}},$$

where n = 6 for YRu$_3$B$_2$ and R = 8.314 J mol$^{-1}$ K$^{-1}$ is the gas constant. The Debye temperature is rather high which is likely caused by a relatively high concentration of the light boron. For another recently described boride superconductor, WB$_{4.2}$, the Debye temperature is even higher reaching 780 K [31].

Fig. 3(c) presents the low-temperature C/T versus T collected in zero field with a large anomaly observed below 0.7 K. Because of the limited temperature range, no clear λ-shaped anomaly, characteristic of a superconducting transition, can be resolved. Nevertheless, having the $\gamma$ value and assuming a weak coupling superconductivity scenario ($\Delta C/\gamma T_c = 1.43$), the expected jump can be estimated, $\Delta C/T_c = 21.5$ mJ mol$^{-1}$ K$^{-2}$. The length of the vertical bar in Figure 3(c) corresponds to this value, enabling an equal entropy construction with a critical temperature of $T_c = 0.63$ K.

To estimate the electron-phonon coupling constant ($\lambda_{el.-ph.}$) we used the inverted McMillan formula [32]:

$$\lambda_{el.-ph.} = \frac{1.04 + \mu^* \ln(\Theta_D/1.45T_C)}{(1-0.62\mu^*)\ln(\Theta_D/1.45T_C) - 1.04}.$$

The usual range of the Coulomb pseudopotential parameter $\mu^*$ is from 0.1 to 0.15. Choosing $\mu^* = 0.13$, with $\Theta_D = 420$ K and $T_c = 0.63$ K this yields $\lambda_{el.-ph.} = 0.35$ and 0.43, respectively, suggesting that YRu$_3$B$_2$ is a weakly-coupled superconductor

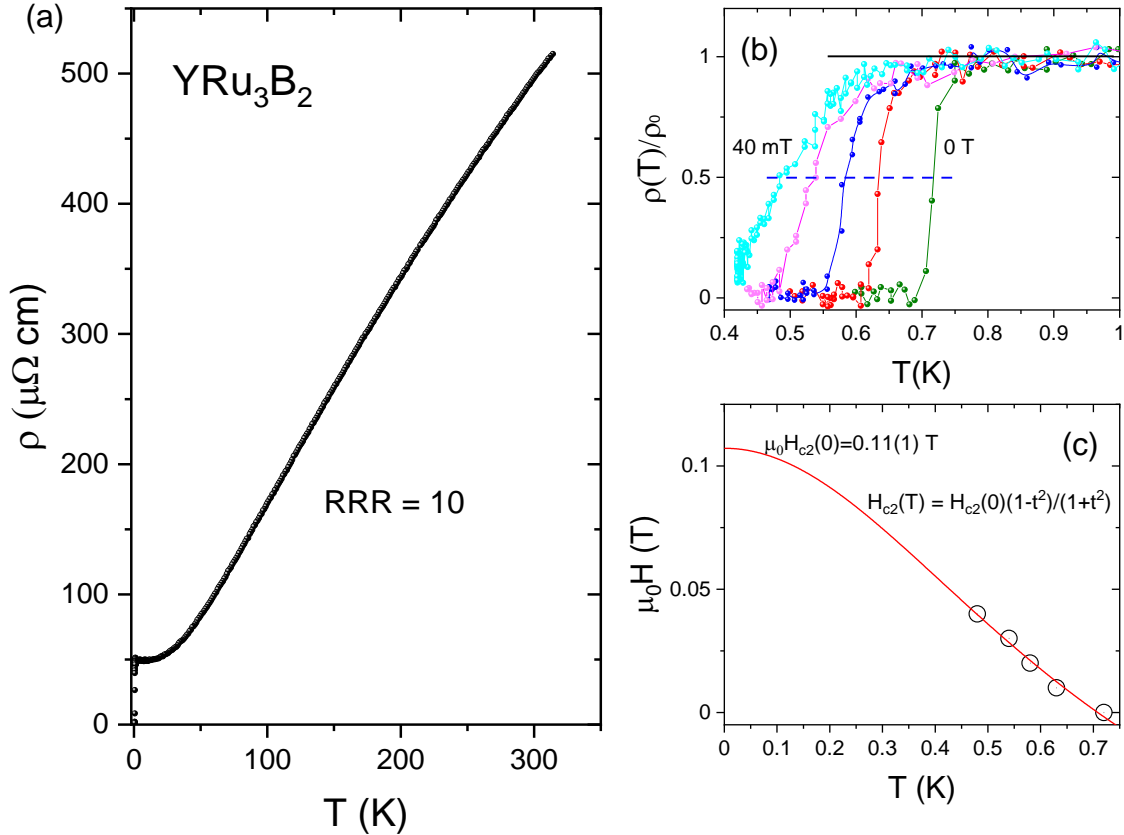

Fig. 4 (a) Temperature-dependent resistivity of YRu$_3$B$_2$ showing a relatively high residual resistivity ratio, RRR = 10. Panel (b) presents the resistivity in the low-temperature range with a superconducting transition visible. Panel (c) shows the temperature dependence of the upper critical field $\mu_0 H_{c2}(T)$, extracted from the resistivity measurements shown in (b), along a fit (red line) used to extrapolate the data to T = 0 K. The upper critical field at 0 K is estimated to be 0.11(1) T.

We followed with electrical resistivity $\rho(T)$ measurements of YRu$_3$B$_2$, presented in Figures 4(a) and (b). As the temperature is decreased, the resistivity decreases, consistent with the metallic character of the compound. At the lowest temperature, the $\rho(T)$ is independent of temperature, and the residual resistivity is approximately $\rho_{res.} \approx 50$ $\mu\Omega$ cm. The calculated residual resistivity ratio, defined as $\rho_{300K}/\rho_{res.}$, is 10, which is rather high value for a polycrystalline material.

Figure 4(b) shows the low temperature $\rho(T)$ in zero magnetic field and in the magnetic fields of 10, 20, 30 and 40 mT. The superconducting critical temperature is defined as the midpoint of the transition to the superconducting state, which is the temperature at which $\rho = 1/2\ \rho_{res.}$ (blue dashed line). From the midpoint criterion, the temperature dependence of the upper critical field is obtained, as shown in Fig. 4(c). Because the observed value of the critical field is rather large for this $T_c$, we assume that $YRu_3B_2$ is a type II superconductor. Hence, the data in Figure 4(c) were fitted with the Ginzburg-Landau formula $\mu_0H_{c2}(T) = \mu_0H_{c2}(0)(1-t^2)/(1+t^2)$, where t is the reduced temperature, defined as t = $T/T_c$. The fit yields $\mu_0H_{c2}(0) = 0.11$ T which is slightly lower than $\mu_0H_{c2}(0)$ reported for elemental Tc and V, which are type II superconductors.

The experimental characterization was complemented with the electronic structure calculations, with the results shown in Fig. 5(a). It is clearly seen that the bands at and around the Fermi energy level, $E_F$, have a dominant Ru 4$d$ character. While no obvious TFB is seen, the bands crossing the Fermi level are relatively weakly dispersive. Symmetry-protected band crossings are observed at the K point of the Brillouin zone, close to the Fermi level. The density of states at the Fermi level is almost completely contributed by Ru 4$d$ states (Fig. 5(b)), highlighting the dominant role of the kagome sublattice in the metallic and superconducting state in $YRu_3B_2$.

The COHP analysis (Fig. 5(c)) reveals that the occupied states exhibit strong Ru–Ru antibonding character, which, consistent with the density-of-states results, dominates the electronic landscape near the Fermi level. In comparison, the Ru–B interactions are only weakly antibonding, and the Y–B interactions weakly bonding. Overall, the COHP results confirm that Ru–Ru interactions play the primary role in governing the conduction in $YRu_3B_2$. As we discussed in our previous articles, the presence of occupied antibonding states appear to correlated with the occurrence of superconductivity in various materials groups [33–35].

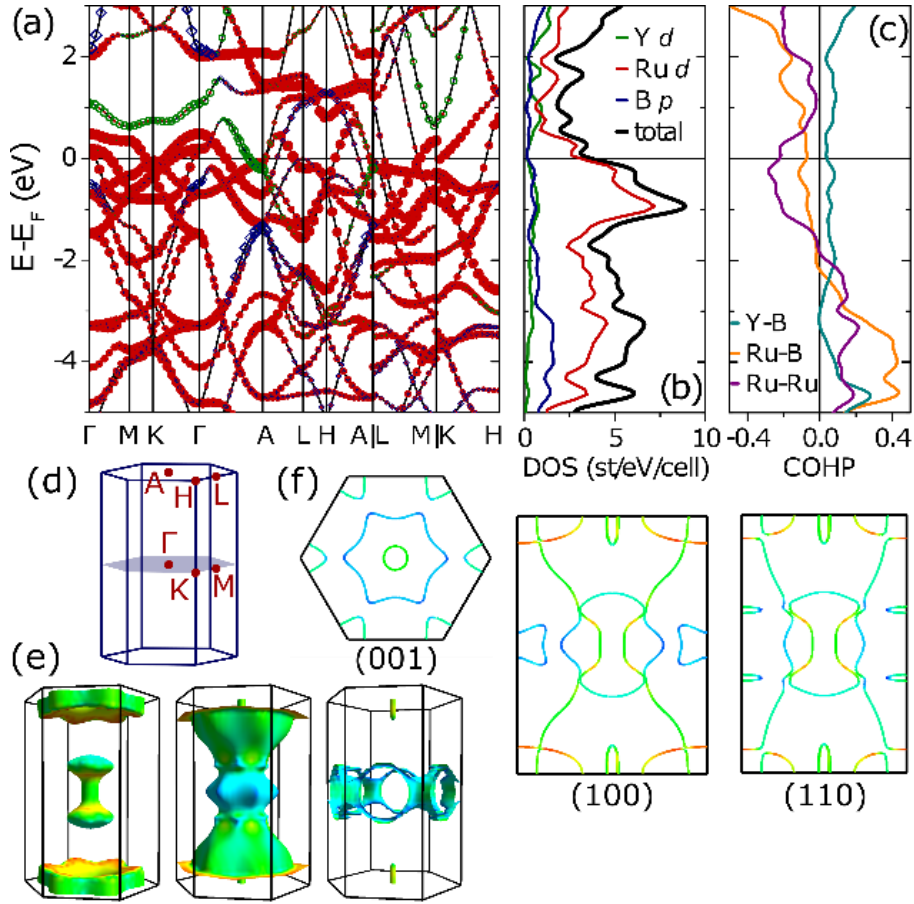

Fig. 5 Band structure (a), density of states (b), and COHP (c) calculated for $YRu_3B_2$. In panel (a) the point size and color shows the relative contribution of atomic orbitals (green – Y 4d, red – Ru 4d, blue – B 2p). The location of high-symmetry points in the first Brillouin zone is shown in panel (d). The three branches of the Fermi surface

(FS) are shown in (e), with the color scale (from blue to red) indicating the relative Fermi velocity. Panel (f) shows three cross-sections through the FS.

Three bands with a dominant Ru 4$d$ character form the Fermi surface (FS) presented in Fig. 5(e,f). The FS branches show both quasi-2D and 3D parts, as seen in Fig. 5(e).

**Conclusions**

Polycrystalline sample of the CeCo$_3$B$_2$-type boride YRu$_3$B$_2$ was successfully synthesized using the arc melting method. Powder x-ray diffraction results confirm the hexagonal space group *P6/mmm* (s.g. # 191) with the lattice parameters: *a* = 5.4826(2) Å and *c* = 3.02166(1) Å. Electrical transport and heat capacity measurements show weak-coupling superconductivity below the critical temperature $T_c$ = 0.63 K (heat capacity) and 0.7 K (resistivity). The results of electronic structure calculations highlight the dominant contribution of Ru 4$d$ orbitals at the kagome layers to the states around the Fermi level, confirming that YRu$_3$B$_2$ is a kagome-based superconductor.

We have also conducted a study of the isoelectronic Sc substitution Y$_{1-x}$Sc$_x$Ru$_3$B$_2$, and as a result, we have determined the Sc solubility limit to be approximately x = 0.15. The chemically doped samples with x = 0.05 and 0.1 have also revealed superconducting behavior with the same $T_c$ = 0.8 K. Conversely, no superconductivity has been observed in a pure YCo$_3$B$_2$ above 0.5 K.

**Acknowledgments**

D.W. acknowledges the Technetium Talent Management Grants program (20/1/2024/IDUB/III.4c/Tc) and D.Y. acknowledges the Nobelium Joining Gdańsk Tech Research Community project (38/2022/IDUB/I.1).